# GDP Trend Deviations and the Yield Spread:
# the Case of Five E.U. Countries

Periklis Gogas* and Ioannis Pragidis

Department of International Economic Relations and Development
Democritus University of Thrace
Komotini 69100, Tel. +306947001079
*Corresponding author, email: pgkogkas@ierd.duth.gr

May 6, 2010

### Abstract

Several studies have established the predictive power of the yield curve in terms of real economic activity. In this paper we use data for a variety of E.U. countries: both EMU (Germany, France, Italy) and non-EMU members (Sweden and the U.K.). The data used range from 1991:Q1 to 2009:Q1. For each country, we extract the long run trend and the cyclical component of real economic activity, while the corresponding interbank interest rates of long and short term maturities are used for the calculation of the country specific yield spreads. We also augment the models tested with non monetary policy variables: the countries' unemployment rates and stock indices. The methodology employed in the effort to forecast real output, is a probit model of the inverse cumulative distribution function of the standard distribution, using several formal forecasting and goodness of fit evaluation tests. The results show that the yield curve augmented with the non-monetary variables has significant forecasting power in terms of real economic activity but the results differ qualitatively between the individual economies examined raising non-trivial policy implications.

**Acknowledgements:** The authors would like to thank the participants at the 68[th] International Atlantic Economic Conference in Boston, 8-11 October 2009, and especially the organizer of the session "Recent Issues in Finance", Prof. Nicholas Apergis.

# 1. Introduction

The yield curve, measuring the difference between short and long term interest rates, has been at the center of recession forecasting. The theoretical justification of this line of work is that since short term interest rates are instruments of monetary policy, and long term interest rates reflect market's expectations on future economic conditions, the difference between short and longer term interest rates may contain useful information to policy makers and other individuals for the corresponding time frame. Furthermore, when the yield curve is upward slopping during recessions, it indicates that there are expectations for future economic upturn. On the other hand, just before recessions, the yield curve flattens or even inverts. De Graeve et al. (2009) explain the predictive power of U.S bond yield curve through demand shocks, wage markup shocks and the investments shocks.

There are two major branches of empirical work in this area: first, simple OLS estimation where researchers try to predict future economic activity and second, probit models that are used to forecast upcoming recessions. The main objective of these two classes of papers is to accommodate the fluctuations of future economic activity taking into account the information that is included in the yield curve and is independent of the exercised monetary policy.

According to the influential paper in this line of research by Estrella and Mishkin (1997), the short end of the yield curve can be affected by the European Central Bank or the Federal Reserve or any other central bank, but the long end will be determined by many other considerations, including long term expectations of inflation and future real economic activity. In their paper, after taking into account monetary policy conducted in four major European countries (France, Germany, Italy and the U.K), Estrella and Mishkin (1997) show that the term structure spread has significant predictive power for both real activity and inflation.

Bonser-Neal and Morley (1997), after examining eleven developed economies, found that the yield spread is a good predictive instrument for future economic activity. In the same vein, Venetis et al. (2003) reached the same conclusion, as did Hamilton and Kim (2002). On the other hand, Kim and Limpaphayom (1997) tested Japan and found evidence that the expected short term interest rate is the only source of predictability for Japan, and not the term premium. Ang et al. (2006), after modeling regressor endogeneity and using data for the period 1952 to 2001, conclude that the short term interest rate has more predictive power than any term spread. They confirm their finding by forecasting GDP out of sample. Bordo et al. (2007) examine the predictive power of the yield curve in U.S.A over the period 1875 to 1997. They find that real growth can be predicted in more accuracy using both the level and slope of the yield curve. In the same vein Chay Fischer et al. (1998) argue that Australian consumption growth can be predicted by the yield curve.

There is also a class of papers that uses probit models to forecast recessions. Wright (2006), using as explanatory variables the Federal Reserve funds rate and the term spread, forecasts recessions six quarters ahead for the U.S. economy. Chauvet and Potter (2001) propose out-of-sample forecasting using standard probabilities and "hitting probabilities" of recession that take into account the length of the business cycle phases. They found that standard probit specification that does not consider the presence of autocorrelated errors and that has time varying parameters due to existence of multiple breakpoints tends to over-predict recession results. In their paper, Estrella et al. (2003) use recent econometric techniques for break-testing to examine whether the empirical relationships are in fact stable. They find that models that predict real activity are somewhat more stable than those that predict inflation, and that binary models are more stable than continuous models.

Feitosa and Tabak (2007), for the case of Brazil, find that the spread possesses information which is not totally explained by the monetary policy.

This paper, following the line of previous work using probit models, concentrates on the predictive power of the yield spread in the context of the European Union. To the best of our knowledge, no such analysis has been done yet for the E.U. Furthermore, as a dependent variable, we use the business cycle instead of the commonly used GDP, and a recession in this paper is defined as a deviation of the business cycle below the trend. We also employ other explanatory variables as well, such as the rate of unemployment and a the corresponding stock exchange indices in an effort to improve the predictive power of the model.

The rest of the paper is organized as follows. In the next section, we describe the data used. We then discuss the methodology and present the empirical results, and finally, in the last section we draw the conclusions for this study.

## 2. The Data

We measure economic activity of five major European countries, Germany, Italy, France, Sweden and the United Kingdom. The data for these countries are quarterly GDP from the OECD data base. They are seasonally adjusted for the period 1994:Q1 to 2009: Q1 for the first three countries and for 1991:Q1 to 2009:Q1 for the rest two. We restrict the analysis to this period as data availability and consistency issues arise for earlier data. Before taking the natural logarithm of the GDP series we apply the OECD seasonally adjusted GDP deflator with 2000 as the base year and we get each country's individual seasonally adjusted real GDP. The aim of the paper is to predict deviations of real output from the long term trend and especially the probability that the GDP of a particular quarter will be below the long run trend. For this reason, we first decompose each country's seasonally adjusted real GDP to the trend and cyclical component employing the Hodrick-Prescott (1997) filter (HP). The HP filter is commonly used in the area of real business cycles. It produces a smooth non-linear trend which is affected more from the long-term fluctuations rather than the short-term ones. The filter's contribution is to distinguish an observed shock into a component that causes permanent effects and a component that has transitory effects on the economy. Furthermore, we address the issue described in the literature of possible bias of the cycle obtained by the HP filter by investigating the robustness of the results to alternative decompositions of the GDP time-series. In doing so, we produce the cyclical component of each country's GDP using alternative specifications for the HP λ parameter i.e. λ=1000, 1600 and 2200. As it is evident from Figure 1, where we illustrate the alternative cyclical components obtained from the three different lambda parameter specifications for the case of Germany, the cyclical component is robust to the alternative values used for λ. This is also the case with the cycles extracted for the other four countries in our sample although we do not provide here the respective Figures[1]. As the qualitative results of the extracted cyclical components are quite similar, we proceed for the rest of the paper using the cycles extracted with the standard lambda parameter for the HP filter for quarterly data, $\lambda = 1600$. Having extracted the cyclical component of each country's real GDP as it is depicted in Figure 2, we then construct the business cycle dummy variable (BS) that takes the value one whenever the cycle is negative implying that the GDP is below trend and the value zero elsewhere. It is important to note here that for the purposes of this paper we define recessions as the negative deviations of GDP from the long term trend. In other words, our aim is to use the yield spread information and other explanatory variables in order to forecast negative values for the cyclical component of the quarterly seasonally adjusted real GDP as it is extracted employing the Hodrick-Prescott (1997) filter. The explanatory variables we use are the corresponding yield spreads and each country's unemployment rate and stock index. All interest rates used in calculating the yield spreads are extracted from the ECB statistics and are the interest rates for the euro area government benchmark bonds with maturities for the long term rate of 1 year, and for the short term rates with maturity of three months. We employ this spread as it is proposed by Chionis et al. (2009) to have the best predictive power. The unemployment rate is obtained from the OECD database. The data for the stock indices are obtained from Six Telekurs. In Table 1, we present a statistical summary of all the explanatory variables.

## 3. Methodology and Empirical Results

We consider 30 alternative models for probit regressions forecasting a quarterly GDP cycle below trend at some point within the next h quarters:

$$Prob(BS_t = 1) = \Phi[\widetilde{\alpha_0} + \widetilde{\alpha_1}(i_{LR,t-i} - i_{SR,t-i})], \ \ i = 1, \ldots \ldots, h \quad (1)$$

where $BS_t$ is the dummy variable that takes the value one every time the cyclical component of the GDP is negative implying a below-trend GDP, and zero elsewhere. $\Phi(\cdot)$ denotes the standard normal cumulative distribution function, $(i_{LR,t-i} - i_{SR,t-i})$ represents the spread between the long and short run interest rates with i = 1,...,6. For the long run interest rates we follow Chionis et al. (2009) and use the 1 year rate, while for the short run rates we use the three months maturity[2]. Finally, $\widetilde{\alpha_0}$ and $\widetilde{\alpha_1}$ are the estimated parameters. Thus, equation 1 is estimated for the aforementioned short and long run interest rates and forecast windows from one to six quarters ahead, a total of 30 probit regressions. The estimated coefficient of the spread is statistically significant at probability 1% for lags 2 through 4 for France, 2 and 3 for Germany, 1 through 3 for Italy and 2 through 6 for Sweden. For the case of the U.K. the spread is significant only for a probability of 10% for lags 2 and 3 and thus the results for the U.K. hereafter must be interpreted with caution. These results are summarized in columns three and four of Table 2. As the main purpose of this paper is the prediction of GDP fluctuations below the long run trend, we formally compare the above significant models for each country in terms of their forecasting ability by calculating the root mean squared error (RMSE), mean absolute error (MAE), and the mean absolute percent error (MAPE) statistics. These statistics are obtained using the following formulas:

$$RMSE = \sqrt{\frac{1}{F}\sum_{f=1}^{F} e_{t+f}^2}$$

$$MAE = \frac{1}{F}\sum_{f=1}^{F} |e_{t+f}|$$

$$MAPE = \frac{1}{F}\sum_{f=1}^{F} \left|\frac{e_{t+f,t}}{y_{t+f}}\right|$$

where $e_{t+f} = y_{t+f} - y^*_{t+f}$, and $y_{t+f}$ is the actual value of the series at period t+f, $y^*_{t+f}$ is the forecast for $y_{t+f}$ and F is the forecast window. Moreover, we report in the last column of Table 2 the McFadden $R^2$ for the probit estimation. According to the statistics at the last four columns of Table 2, we select a forecast window of three quarters for France, Germany and the U.K., two quarters for Italy and six for Sweden. The range of values for the McFadden $R^2$ between 0.136 and 0.310 (with the exception of the U.K.) is considered a good fit as this statistic tends to be smaller than standard $R^2$.

---

[1] Of course these are available from the authors upon request.
[2] See Chionis, Gogas and Pragidis (2009) for an explanation for this selection.

Next, in an effort to examine whether other variables from the real economy can add any informational content to the forecasts of GDP, we estimate the following probit regressions:

$$Prob(BS_t = 1) = \Phi[\widetilde{\alpha_0} + \widetilde{\alpha_1}(i_{LR,t-i} - i_{SR,t-i})] + \tilde{a}_u u_{t\_1} \quad (2)$$

$$Prob(BS_t = 1) = \Phi[\widetilde{\alpha_0} + \widetilde{\alpha_1}(i_{LR,t-i} - i_{SR,t-i})] + \tilde{a}_s s_{t\_1} \quad (3)$$

where $u_t$ is the unemployment rate for each country and $s_t$ is the stock market index of the respective country, and $\widetilde{\alpha_u}$, $\widetilde{\alpha_s}$ are their estimated coefficients. The significance of the inclusion of the non-monetary variables is confirmed by a Wald test where we test the join hypothesis that the coefficients of unemployment and the stock index are equal to zero: $\tilde{a}_u = \tilde{a}_s = 0$. We reject the null hypothesis for all countries with the exception of France - the results are reported in Table 3. The augmenting non-monetary variables appear to improve significantly both the explanatory power and the forecasting ability of the models as we can see in Table 4 where we present the three forecasting criteria and the McFadden $R^2$ statistic. Thus, we employ these models in the effort to forecast a below-trend real GDP for the five countries. The results of these forecasts are presented in Figure 3 where we graph the forecasted probability of a recession along with the seasonally adjusted real GDP cyclical component of each country. As it can be seen in Figure 3, the predictive power of the estimated model in terms of the forecasted probabilities of the studied countries' GDP deviations from the trend is very high. It seems that the yield spread between the 1 year and the three month interest rates augmented by the unemployment rate and the corresponding stock index is a very good predictor of the cyclical behavior of GDP in terms of its deviations from the long run trend.

## 4. Conclusion

In this paper, we have used several probit models to examine the forecasting power of the yield spread between long term and short term interest rates in terms of real GDP deviations from the long-run trend. Five E.U. countries were studied- France, Germany, Italy, Sweden and the U.K. Moreover, we augmented the estimation models with other non-monetary variables, the unemployment rate and the stock markets indices of the countries in question, as they appeared significant in adding explanatory and forecasting power to our basic yield-spread models. Overall, the final model used for forecasting appears very efficient to forecast deviations of real output from the long run trend according to the standard formal goodness of fit tests employed. The results of course generate obvious policy implications. The policymaker can use the information provided by the yield spread, unemployment and the stock market today in order to estimate the probability of obtaining a below-trend real output two to six quarters ahead. A shrinking yield spread or in other words a yield curve with a diminishing slope in the short rates domain may be the signal for an upcoming below-trend real output. Thus, the policymaker who is concerned with stable growth and targets small fluctuations of real GDP—especially downwards—can use this information and loosen monetary policy in an effort to reduce short-term interest rates (directly affected by monetary policy), increase the spread, and lower the probability of a below-trend real GDP. In this manner, successful intervention in the term structure of interest rates could shorten the below-trend cycle and/or make the fluctuation milder.

**Table 1.** Statistical Summary of the Explanatory Variables

Panel A: Monetary Data

| | 1-Year Interest Rate | | | | | 3-Month Interest Rate | | | | |
|---|---|---|---|---|---|---|---|---|---|---|
| | France | Germany | Italy | Switzerland | UK | France | Germany | Italy | Switzerland | U... |
| Mean | 4.07 | 4.07 | 4.07 | 3.92 | 5.66 | 3.99 | 3.99 | 3.99 | 4.34 | 5... |
| Median | 4.09 | 4.09 | 4.09 | 4.06 | 5.74 | 4.06 | 4.06 | 4.06 | 4.12 | 5... |
| Maximum | 7.73 | 7.73 | 7.73 | 6.34 | 7.77 | 7.14 | 7.14 | 7.14 | 9.16 | 7... |
| Minimum | 2.14 | 2.14 | 2.14 | 2.14 | 2.37 | 2.01 | 2.01 | 2.01 | 1.63 | 2... |
| Std. Dev. | 1.41 | 1.41 | 1.41 | 1.21 | 1.19 | 1.43 | 1.43 | 1.43 | 1.91 | 1... |
| Skewness | 0.62 | 0.62 | 0.62 | 0.13 | -0.21 | 0.47 | 0.47 | 0.47 | 1.03 | -0... |
| Kurtosis | 2.90 | 2.90 | 2.90 | 1.99 | 2.63 | 2.45 | 2.45 | 2.45 | 3.47 | 2... |
| Jarque-Bera | 3.98 | 3.98 | 3.98 | 2.77 | 0.78 | 3.00 | 3.00 | 3.00 | 11.24 | 0... |
| Probability | 0.14 | 0.14 | 0.14 | 0.25 | 0.68 | 0.22 | 0.22 | 0.22 | 0.00 | 0... |
| Observations | 61 | 61 | 61 | 61 | 61 | 61 | 61 | 61 | 61 | |

Panel B: Non Monetary Data

| | Unemployment Rate | | | | | Stock Index (in logs) | | | | |
|---|---|---|---|---|---|---|---|---|---|---|
| | France | Germany | Italy | Switzerland | UK | France | Germany | Italy | Switzerland | U... |
| Mean | 9.11 | 8.87 | 9.18 | 7.10 | 6.07 | 8.20 | 8.36 | 9.88 | 6.51 | 8... |
| Median | 8.90 | 8.60 | 8.90 | 6.63 | 5.47 | 8.23 | 8.47 | 9.99 | 6.57 | 8... |
| Maximum | 10.90 | 11.40 | 11.40 | 10.30 | 9.77 | 8.78 | 8.99 | 10.42 | 7.22 | 8... |
| Minimum | 7.20 | 6.90 | 6.10 | 4.77 | 4.63 | 7.50 | 7.61 | 9.13 | 5.63 | 8... |
| Std. Dev. | 1.12 | 1.09 | 1.82 | 1.66 | 1.41 | 0.37 | 0.40 | 0.40 | 0.44 | 0... |
| Skewness | 0.06 | 0.48 | -0.21 | 0.36 | 1.13 | -0.43 | -0.43 | -0.56 | -0.49 | -0... |
| Kurtosis | 1.63 | 2.49 | 1.58 | 1.89 | 3.07 | 2.11 | 2.07 | 2.01 | 2.39 | 2... |
| Jarque-Bera | 4.80 | 3.02 | 5.61 | 4.47 | 12.95 | 3.85 | 4.09 | 5.70 | 3.34 | 4... |
| Probability | 0.09 | 0.22 | 0.06 | 0.11 | 0.00 | 0.15 | 0.13 | 0.06 | 0.19 | 0... |
| Observations | 61 | 61 | 61 | 61 | 61 | 61 | 61 | 61 | 61 | 6... |

Table 2. Forecasting Model Selection Criteria

| Predicting Spread | | | | Forecasting criteria | | | |
|---|---|---|---|---|---|---|---|
| Country | Spread | Forecast Window | Prob. | RMSE | MAE | MAPE | McFadden R2 |
| France | 1y-3m | 2 Qrts | 0.00 | 0.427 | 0.365 | 18.280 | 0.218 |
| | 1y-3m | 3 Qrts * | 0.00 | 0.404 | 0.329 | 16.605 | 0.284 |
| | 1y-3m | 4 Qrts | 0.00 | 0.446 | 0.402 | 20.468 | 0.151 |
| Germany | 1y-3m | 2 Qrts | 0.00 | 0.455 | 0.414 | 20.966 | 0.130 |
| | 1y-3m | 3 Qrts * | 0.00 | 0.454 | 0.412 | 20.982 | 0.136 |
| Italy | 1y-3m | 1 Qrts | 0.00 | 0.460 | 0.426 | 21.527 | 0.114 |
| | 1y-3m | 2 Qrts * | 0.00 | 0.458 | 0.424 | 21.426 | 0.116 |
| | 1y-3m | 3 Qrts | 0.00 | 0.465 | 0.435 | 21.926 | 0.097 |
| Sweden | 1y-3m | 2 Qrts | 0.00 | 0.462 | 0.426 | 21.375 | 0.108 |
| | 1y-3m | 3 Qrts | 0.00 | 0.455 | 0.416 | 20.968 | 0.125 |
| | 1y-3m | 4 Qrts | 0.00 | 0.431 | 0.377 | 19.133 | 0.187 |
| | 1y-3m | 5 Qrts | 0.00 | 0.408 | 0.345 | 17.597 | 0.237 |
| | 1y-3m | 6 Qrts * | 0.00 | 0.379 | 0.303 | 15.418 | 0.310 |
| U.K. | 1y-3m | 2 Qrts | 0.07 | 0.487 | 0.475 | 23.820 | 0.036 |
| | 1y-3m | 3 Qrts * | 0.05 | 0.486 | 0.471 | 23.557 | 0.043 |

An asterisk denotes the selected forecast window.

Table 3. Hypothesis Testing

| Probability of Hypothesis: | $\tilde{a}_u = \tilde{a}_s = 0$ | |
|---|---|---|
| Country | F-Stat | $X^2$ |
| France | 0.181 | 0.171 |
| Germany | 0.012 | 0.008 |
| Italy | 0.057 | 0.049 |
| Sweden | 0.045 | 0.039 |
| U.K. | 0.007 | 0.004 |

Table 4. Forecasting Model Selection Criteria with non-Monetary Variables

| Predicting Spread | | | Forecasting criteria | | | |
|---|---|---|---|---|---|---|
| Country | Spread | Forecast Window | RMSE | MAE | MAPE | McFadden R2 |
| France | 1y-3m | 3 Qrts | 0.39 | 0.31 | 15.64 | 0.33 |
| Germany | 1y-3m | 3 Qrts | 0.41 | 0.34 | 16.68 | 0.28 |
| Italy | 1y-3m | 2 Qrts | 0.43 | 0.38 | 19.43 | 0.19 |
| Sweden | 1y-3m | 6 Qrts | 0.35 | 0.26 | 13.41 | 0.39 |
| U.K. | 1y-3m | 3 Qrts | 0.45 | 0.39 | 19.73 | 0.18 |

**Figure 1.** Cyclical component sensitivity to alternative parameter specifications for the case of Germany.

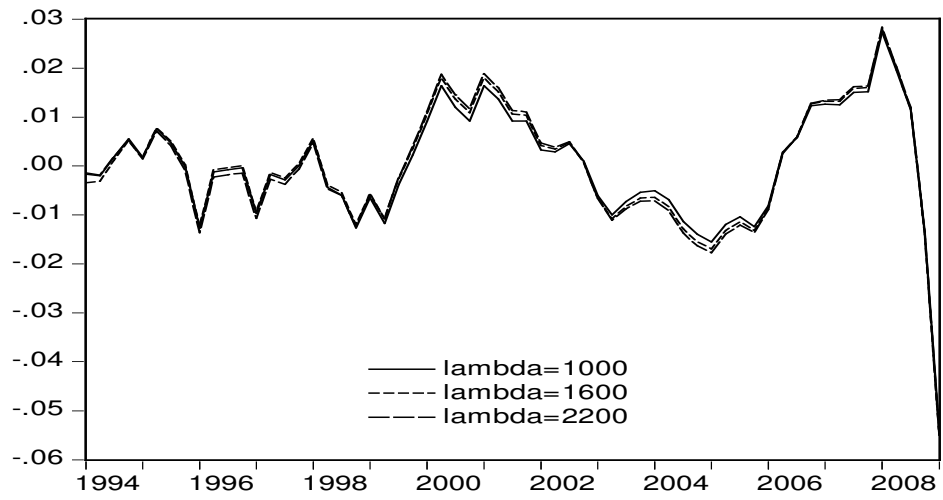

**Figure 2.** Extracted cyclical components of long run real GDP.

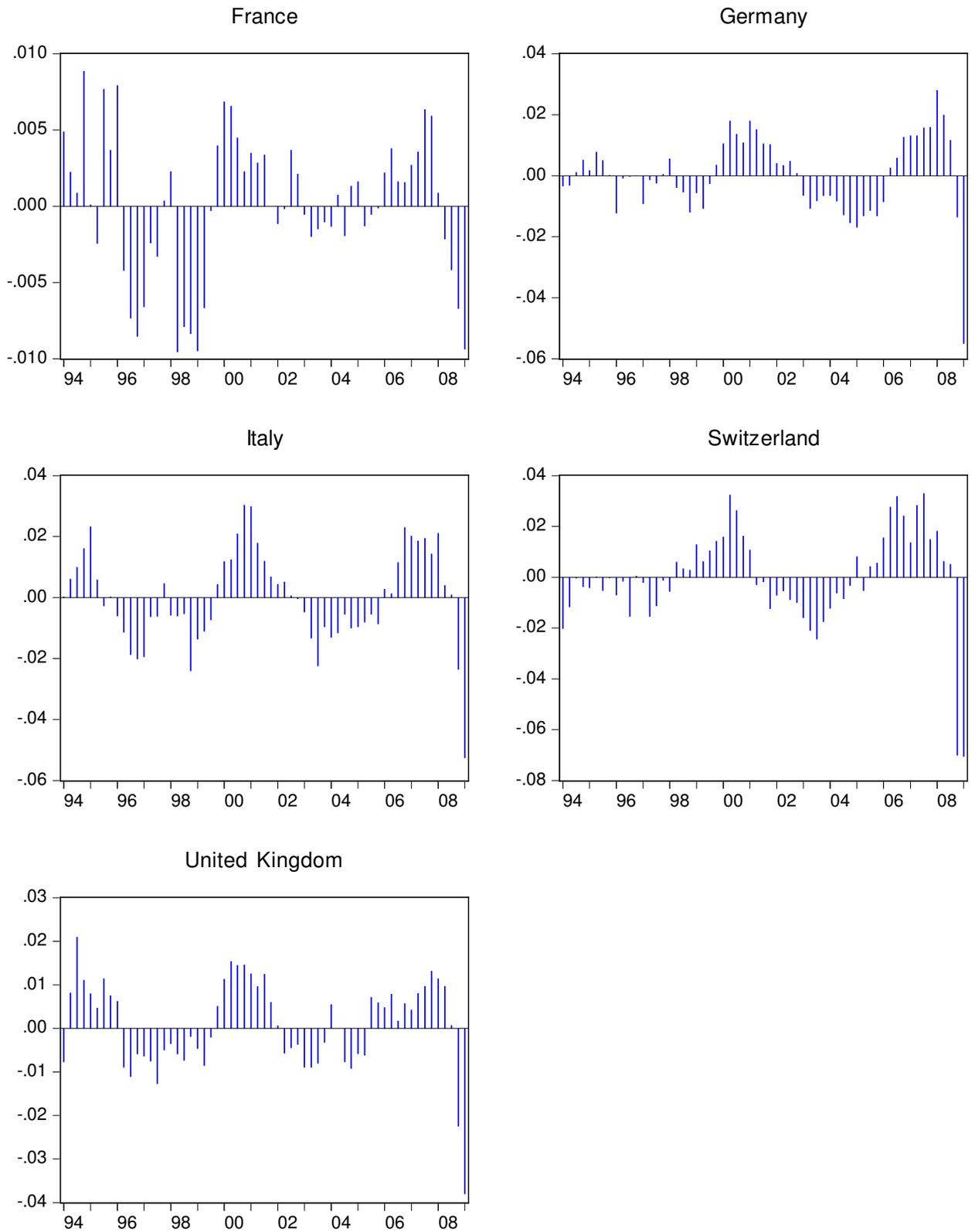

**Figure 3.** GDP Cyclical Component and Forecasted Probability

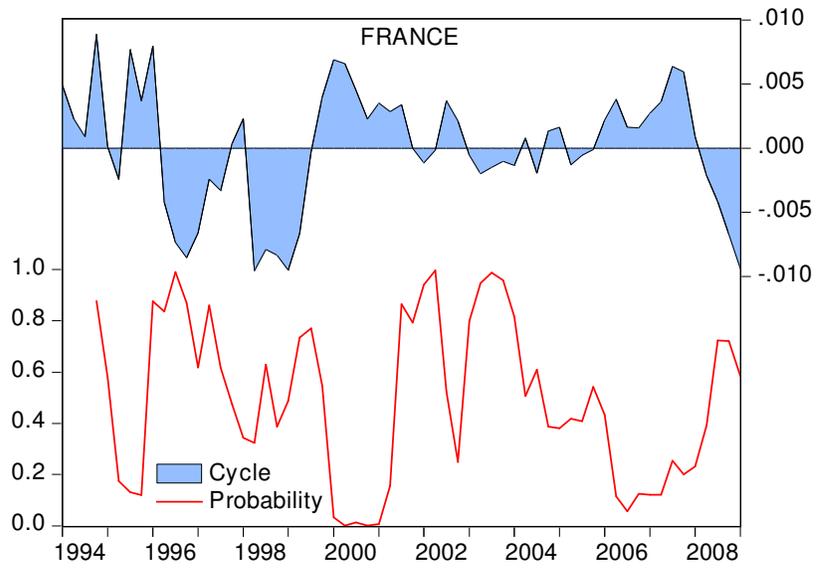

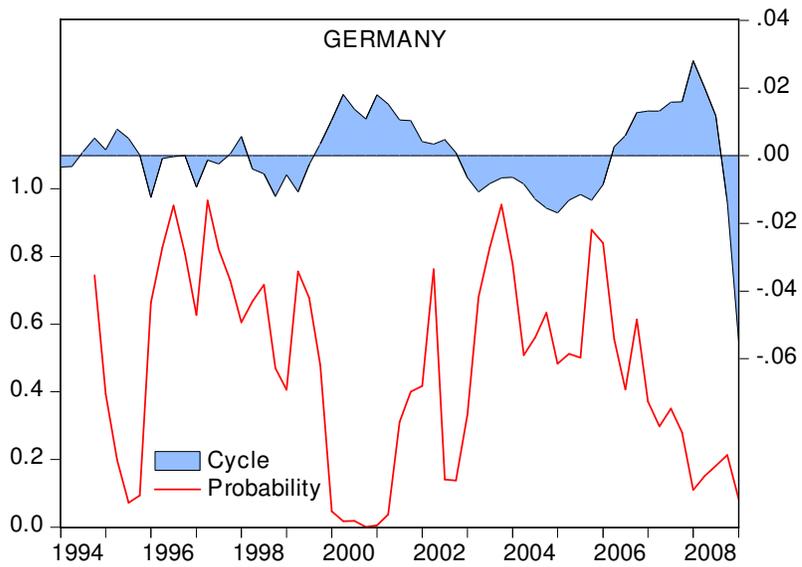

**Figure 3 (Continued).** GDP Cyclical Component and Forecasted Probability

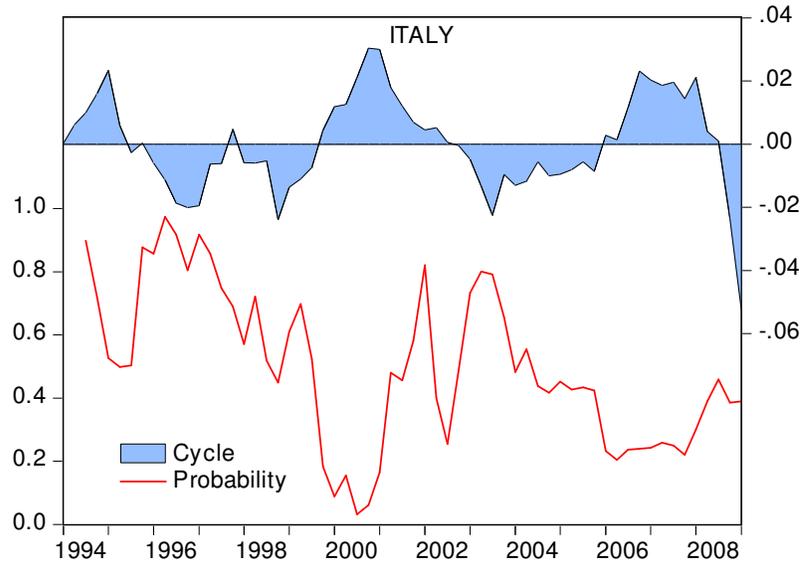

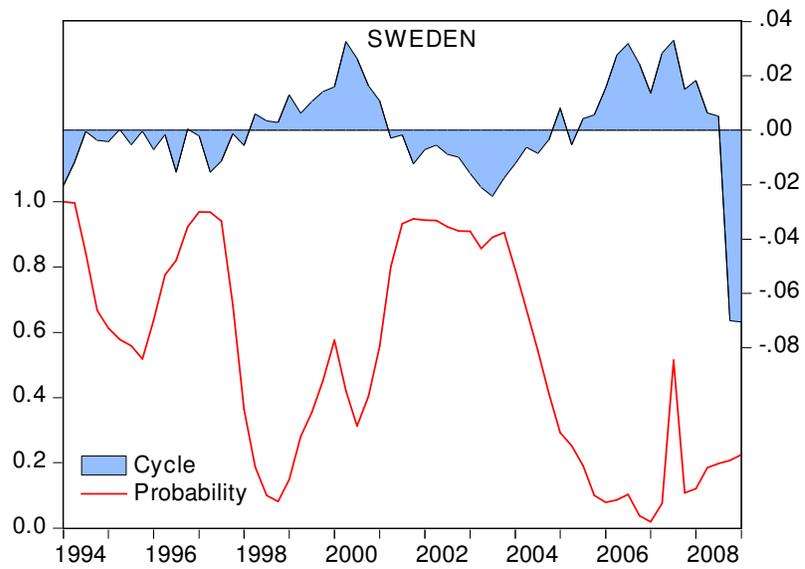

**Figure 3 (Continued).** GDP Cyclical Component and Forecasted Probability

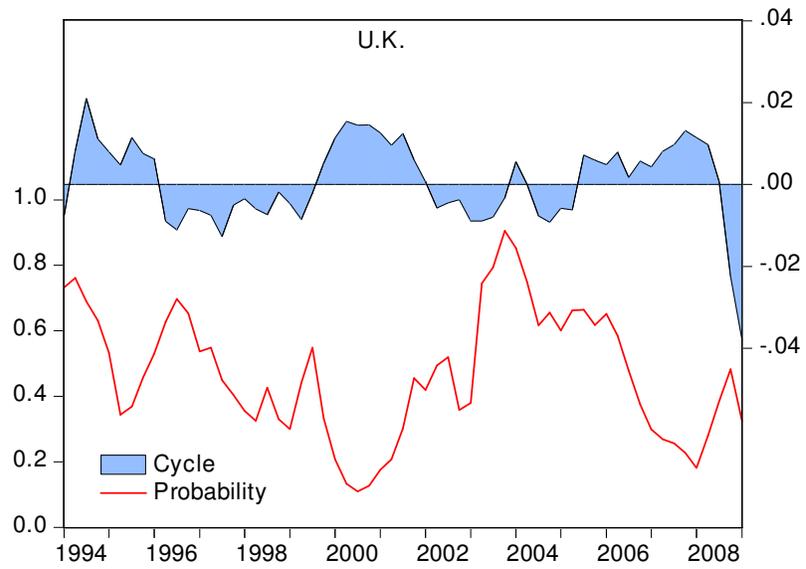